\documentclass[prl,aps,amsmath,amssymb,amsfonts,twocolumn,footinbib]{revtex4-1}

\usepackage[german,american,english]{babel}
\usepackage{graphicx}
\usepackage{dcolumn}
\usepackage{multirow}
\usepackage{bm}
\usepackage{latexsym}
\usepackage{layout}
\usepackage{verbatim}
\usepackage{xcolor}
\usepackage[T1]{fontenc}
\usepackage[caption=false]{subfig}

\newcommand{\bea}{\begin{eqnarray*}}
\newcommand{\eea}{\end{eqnarray*}}
\newcommand{\bne}{\begin{equation*}}
\newcommand{\ede}{\end{equation*}}

\newcommand{\bnen}{\begin{equation}}
\newcommand{\eden}{\end{equation}}
\newcommand{\bean}{\begin{eqnarray}}
\newcommand{\eean}{\end{eqnarray}}
\newcommand{\bsen}{\begin{subequations}}
\newcommand{\esen}{\end{subequations}}
\newcommand{\ba}{\arraycolsep 0.3ex \begin{array}{rl}}
\newcommand{\ea}{\end{array}}

\newcommand{\bna}{\begin{array}}
\newcommand{\eda}{\end{array}}
\newcommand{\bnm}{\begin{enumerate}}
\newcommand{\edm}{\end{enumerate}}

\newcommand {\dbkt} [2] {\langle #1 | #2 \rangle}

\newcommand {\pd} [2] {\frac{\partial #1}{\partial #2}}

\def\hh{\hskip 0.1em}

\def\ket#1{{\big| \hh #1\hh \big\rangle}}

\begin{document}

\title{Zitterbewegung velocity in semiclassical electron dynamics}

\author{Dimitrie Culcer}
\affiliation{School of Physics, The University of New South Wales, Sydney 2052, Australia}

\begin{abstract}
Zitterbewegung plays a major role in electron dynamics in solids, yet is not captured in conventional semiclassical treatments. Here, starting from the quantum Liouville equation, I identify a new \textit{Zitterbewegung velocity}, which involves the symmetric and antisymmetric components of the quantum geometric tensor oscillating out of phase. The Zitterbewegung velocity resolves the \textit{position-shift paradox}, recovering the field-induced shift in an electron's position by integrating the semiclassical equations, and is directly related to the famous \textit{minimum conductivity} of massless Dirac fermions.
\end{abstract}
\date{\today}
\maketitle


\textit{Introduction}. The semiclassical description of carrier dynamics has long served as a cornerstone of transport theory in solids \cite{Chambers1966, Ashcroft76, GaneshSundaram1999}. Extensively developed in the context of anomalous transport, where Berry-phase effects give rise to intrinsic Hall responses and related phenomena \cite{Jungwirth2002, Sinitsyn_2007JPCM, AHE-RevModPhys.82.1539, Matsumoto2011, ChengRan2016, GaoYang2019, Zhang2019, Culcer_AHE_2024}, the semiclassical model is nowadays routinely used to describe transport in systems with nontrivial topological structures \cite{SHINDOU2005,Dumitrescu2012,Son2013,SongJustin2015,YangShengyuan2015,Lensky2015,ChangMingChe2015,Gorbar2018,Misaki2018,Araki2018,Alexandradinata2018,DasKamal2019,Nandy2019,Hou2019,ZhuPenghao2021,Yokoyama2021,WangZhi2021,DasKamal2021,Chaudhary2021,Bhowal2021, Battiato2021, Raimondi_EPL2023}, including nonlinear electromagnetic responses \cite{Moore2010, Sodemann2015, Morimoto2016, ZhenChuanchang2019, Golub2020}, and has emerged as a central tool in computational and first-principles studies of nonequilibrium phenomena in multiband systems \cite{Wang2006, Wang2007, Gradhand_2012, HeYan2012, ChenHua2014, Olsen2015, WanxiangFeng2016, XinDai2017, Martiny2019, Wuttke2019, DuShiqiao2020, KTLaw2020, HeWen2021}. A major strength is its ability to incorporate topological effects while maintaining a transparent separation between Fermi-surface and Fermi-sea contributions to transport \cite{Chang_2008,Interband-Coherence-PRB-2017-Dimi,Stedman_2019,Stedman2020}.

Despite this success, the semiclassical formulation does not capture all aspects of interband dynamics. One well-known area of disagreement between semiclassical and quantum predictions comprises disordered systems, since the semiclassical model conventionally describes dynamics between collisions. This gap has been partially remedied by incorporating the side-jump velocity into the semiclassical equations \cite{JE-PRR-Rhonald-2022}. Yet recent years have highlighted another inconsistency present even in disorder-free systems, which may be termed the \textit{position-shift paradox}. Quantum mechanics predicts a position shift of a Bloch-electron wave packet in an electric field. Semiclassically, this shift appears when considering an electric-field-induced correction to the basis functions, leading to a change in the expectation value of a wave packet's center of mass \cite{GaoYangNiu2014}. However, the steady-state position shift cannot be found by straightforwardly integrating the conventional semiclassical equation of motion for the velocity. This is paradoxical within the semiclassical framework, where the equations of motion are intended to provide a closed description of quasiparticle dynamics, and one expects to derive all field-induced corrections directly from the equations of motion \cite{GaneshSundaram1999,Chang_2008,XiaoDi2010}.

In this work I argue that this paradox stems from the failure of the semiclassical model, in its conventional formulation, to account for Zitterbewegung. In a solid-state context Zitterbewegung refers to interband oscillatory motion which appears prominently in nonequilibrium dynamics in external fields \cite{Zawadzki_Zitterbewegung_PRB2005, Schliemann2006, Culcer2006, Katsnelson_Zitterbewegung_EPJB2006, WinklerZuelicke2007, RusinZawadzki2009, Zawadzki2010, Hamner2014}. It is usually perceived as an inherently quantum-mechanical effect without an analog in semiclassical dynamics. Here, by deriving the semiclassical equations directly from the quantum Liouville equation for the density matrix \cite{Luttinger_AHE_PR58, Vasko2005, Interband-Coherence-PRB-2017-Dimi, JE-PRR-Rhonald-2022}, I show that this shortcoming can be remedied by incorporating a new, intrinsic, time-dependent term into the semiclassical equations of motion in an electric field, which I term the \textit{Zitterbewegung velocity}. To recover this term in the velocity one needs to integrate the nonequilibrium quantum Liouville equation from a fixed starting time up to the observation time, relinquishing the customary steady-state linear-response assumption that places the initial time of the electric field in the infinitely distant past. The central result of this work is the modified semiclassical equation for the electron velocity in a clean system
\begin{equation}
    \dot{\bm r}_m = \displaystyle \frac{1}{\hbar}\pd{\varepsilon_m}{\bm k} + \frac{e}{\hbar} \, {\bm E} \times {\bm \Omega}_m + {\bm \zeta}_m (t),
\end{equation}
where $\varepsilon_m$ is the energy of band $m$, ${\bm \Omega}_m$ the Berry curvature, and the \textit{Zitterbewegung velocity} ${\bm \zeta}_m$ takes the form
\begin{equation}\label{vzb}
\arraycolsep 0.3ex
\begin{array}{rl}
\displaystyle \zeta^i_m (t) = & \displaystyle \frac{ieE_j}{\hbar} \sum_{n \ne m} {\mathcal R}^i_{mn} {\mathcal R}^j_{nm} \, e^{- i \omega_{mn} t} + h.c.
\end{array}
\end{equation}
Here $h.c.$ represents the Hermitian conjugate, the frequency $\omega_{mn} = (\varepsilon_m - \varepsilon_n)/\hbar$, and the Berry connection ${\mathcal R}^i_{mn} = \dbkt{u_{m{\bm k}}}{i\partial_{k_i}u_{n{\bm k}}}$, with $\ket{u_{m{\bm k}}}$ the lattice-periodic part of the Bloch wave function \cite{GaneshSundaram1999,Chang_2008,XiaoDi2010}.

The Zitterbewegung velocity involves both the symmetric and antisymmetric components of the quantum geometric tensor oscillating out of phase, with a frequency determined by the energy differences between pairs of bands. Importantly, when the Zitterbewegung velocity is integrated over time to find the change in the electron position, it immediately yields the field-induced position shift, which is independent of time in the long-time limit \cite{GaoYangNiu2014}. In contrast, the expectation value of the macroscopic charge current is largely unaffected by the oscillatory Zitterbewegung terms, since integration over phase space generally causes them to dephase. This is demonstrated below for several models. A notable exception is the case of massless Dirac fermions as in graphene and topological-insulator surface states, where under certain circumstances the Zitterbewegung velocity yields a correction to the expectation value of the charge current. This correction leads directly to the minimum conductivity of massless Dirac fermions at the Dirac point \cite{Sinitsyn2006_II,Tworzydlo2006}, and the revised semiclassical procedure introduced in this work captures this inherently quantum-mechanical phenomenon as well. Importantly, while interband coherence and oscillatory contributions to observables are well known in density-matrix and Kubo formulations, their identification as an explicit contribution to the semiclassical equation of motion has not been previously recognized. In the present formulation, the Zitterbewegung velocity restores closure of the semiclassical dynamics by ensuring that the electron position follows directly from time integration of the velocity'

\textit{Methodology}. Following the density-matrix program developed recently for interband coherence and for disordered systems \cite{Interband-Coherence-PRB-2017-Dimi, JE-PRR-Rhonald-2022}, I derive the semiclassical equations in an electric field using the density matrix. I focus on a disorder-free, noninteracting electron system, assume zero temperature, and restrict attention to nondegenerate bands. The system is described by a single-particle density operator $\hat \rho$ and the expectation value of the velocity component $\hat{v}_i$ is given by $\mathrm{Tr}(\hat {v}_i \hat \rho)$, where Tr denotes the operator trace. One begins by implementing linear-response theory in an external electric field. Having found the nonequilibrium correction to the density matrix, one takes its trace with the velocity operator, which in the crystal-momentum representation has both intraband and interband contributions. In the intrinsic case considered here only the interband velocity is relevant, as it encapsulates interband dynamics leading to Zitterbewegung. Next, I show that the nonequilibrium correction to the velocity expectation value can be recast in terms of the equilibrium Fermi-Dirac function for a single band, and I argue that the term multiplying the Fermi-Dirac distribution should be interpreted as the change in the velocity of a single electron at wave vector ${\bm k}$ induced by the electric field. In addition to the well-known group-velocity and Berry-curvature terms, which emerge trivially, the electron velocity contains an oscillatory term due to Zitterbewegung.

We begin with the density operator $\hat \rho$. The matrix elements of this operator form the density matrix, which is expressed here in the crystal-momentum representation spanned by the Bloch wave functions $\ket{\psi_{m{\bm k}}} = e^{i{\bm k}\cdot{\bm r}}\ket{u_{m{\bm k}}}$, with $\ket{u_{m{\bm k}}}$ the lattice-periodic part introduced above. The $\ket{\psi_{m{\bm k}}}$ are eigenstates of the crystal Hamiltonian $H_0$. The equilibrium density matrix $\rho_0$ is diagonal in the crystal-momentum representation, with matrix elements $\rho_{0;mn} = f_m \delta_{mn}$, where $f_m \equiv f(\varepsilon_m)$ is the Fermi-Dirac distribution. Throughout the paper, wherever possible, I suppress the wave-vector indices, with the understanding that $\varepsilon_m$, $f_m$, and $\rho$ are functions of ${\bm k}$. A constant, uniform electric field is encapsulated in the potential $V = e{\bm E}\cdot{\bm r}$, where ${\bm r}$ refers to the position operator. The total Hamiltonian is thus $H = H_0 + V$. The focus of the paper is the velocity, since the acceleration $\hbar\dot{\bm k} = - e{\bm E}$ emerges immediately from the commutator of the crystal momentum with $V$ \cite{GaneshSundaram1999,XiaoDi2010}.

When an electric field is applied the density matrix takes the form $\rho = \rho_0 + \rho_E$. To determine the nonequilibrium correction to the density matrix $\rho_E$ to linear order in ${\bm E}$ we solve
\begin{equation}
    \pd{\rho_E^{mn}}{t} + i \, \omega_{mn} \, \rho_{E;mn} = \frac{ieE_j}{\hbar} {\mathcal R}^j_{mn} (f_m - f_n),
\end{equation}
which is the clean-limit interband-coherence equation familiar from the density-matrix formulation of electric-field response \cite{Interband-Coherence-PRB-2017-Dimi}. To recover Zitterbewegung at the microscopic level we need to extricate ourselves from the steady-state viewpoint by retaining the explicit time dependence of the density matrix. Hence we do \textit{not} choose to turn on the electric field in the infinitely far past and we do not resort to adiabatic regularization factors in the time integral. Instead, we turn on the field $t = 0$ and integrate up to the observation time $t$. One could account for this by writing ${\bm E}(t) = {\bm E}\Theta(t)$. This gives
\begin{equation}
\arraycolsep 0.3ex
\begin{array}{rl}
\displaystyle \rho_E^{mn} (t) = & \displaystyle e E_j {\mathcal R}^j_{mn{\bm k}} \frac{f^m_{\bm k} - f^n_{\bm k}}{\varepsilon^m_{\bm k} - \varepsilon^n_{\bm k}} \, \bigl(1 - e^{- i\omega_{mn{\bm k}} t}\bigr).
\end{array}
\end{equation}

\textit{Zitterbewegung velocity}. To identify the Zitterbewegung contribution to the electron velocity, we trace the nonequilibrium density matrix with the velocity operator. The velocity operator has matrix elements $v^i_{mn} = (1/\hbar) \, [\delta_{mn} \, \partial{\varepsilon_m}/\partial k_i + i{\mathcal R}^i_{mn} (\varepsilon_m - \varepsilon_n)]$ \cite{GaneshSundaram1999,Chang_2008,XiaoDi2010}. Since $\rho_E$ in the intrinsic case has only off-diagonal elements, only the interband velocity is relevant. Tracing $\rho_E$ with the interband velocity results in
\begin{equation}
\arraycolsep 0.3ex
\begin{array}{rl}
\displaystyle \sum_{mn} v^i_{nm} \rho_{E;mn} = & \displaystyle - ieE_j \sum_{mn}\{ \bigl({\mathcal R}^i_{{nm}} {\mathcal R}^j_{mn} - h.c.\bigr) \, f_m \\ [2ex]
& \displaystyle - \bigl({\mathcal R}^i_{nm} {\mathcal R}^j_{mn} \,  e^{- i \omega_{mn} t} - h.c.\bigr) \, f_m \},
\end{array}
\end{equation}
where $h.c.$ represents the Hermitian conjugate. The time-independent term leads immediately to the usual Berry-curvature anomalous velocity, with the usual sign and magnitude \cite{GaneshSundaram1999,XiaoDi2010}, and will not be considered further. When the time-dependent term on the second line is expanded it is immediately seen to take the form $\sum_m \zeta^i_m(t) f_m$, with $\zeta^i_m(t)$ given by Eq.~\eqref{vzb}. \footnote{The choice of initial time is arbitrary and I have taken it to be zero. Replacing that with a finite initial time $t_0$ simply changes $t$ to $t-t_0$ in the oscillatory factors, for example $\sin \omega (t-t_0)$, without changing the physical conclusions.}

One implication of the Zitterbewegung velocity is that an electron in one band feels the influence of adjacent bands mediated by the gauge structure encoded in the Berry connection. It is enlightening to separate Eq.~\eqref{vzb} into sinusoidal and cosinusoidal parts. A short manipulation reveals that ${\bm \zeta}_m(t) = \sum_n {\bm c}_{mn} \cos \omega_{mn}t + {\bm s}_{mn} \sin \omega_{mn}t$, where the cosinusoidal part corresponds to the antisymmetric combination $c_{mn}^i = ieE_j ({\mathcal R}^i_{{nm}} {\mathcal R}^j_{mn} - {\mathcal R}^j_{{nm}} {\mathcal R}^i_{mn})$ and the sinusoidal part is the symmetric combination $s_{mn}^i = eE_j ({\mathcal R}^i_{{nm}} {\mathcal R}^j_{mn} + {\mathcal R}^j_{{nm}} {\mathcal R}^i_{mn})$. Thus the Zitterbewegung correction involves precisely the structures that enter the quantum geometric tensor: its antisymmetric Berry-curvature sector and its symmetric quantum-metric sector \cite{Provost1980, GaoYangNiu2014}. However, the terms entering the Zitterbewegung velocity are neither the Berry curvature nor the quantum metric by themselves; rather, each interband contribution is weighted by an oscillatory function of the energy difference between the corresponding pair of bands.

\textit{Field-induced position shift}. When we integrate the Zitterbewegung velocity over time in order to determine the change in position, the oscillatory terms yield $\int_0^t dt' \, \cos\omega_{mn} t' = \sin(\omega_{mn} t)/\omega_{mn}$ and $\int_0^t dt' \, \sin \omega_{mn} t' = [1 - \cos(\omega_{mn} t)]/\omega_{mn}$. In the steady state the oscillatory pieces dephase, but the time-independent pieces survive. Hence the $\sin$ term in the velocity, which one might be tempted to dismiss as transient, generates a constant contribution once integrated over time. Indeed, integrating $s_{mn}^i \sin \omega_{mn} t$ gives the change in position
\begin{equation}
    \Delta r^i_{E; m} = \sum_{n\neq m}\frac{eE_j\bigl({\mathcal R}^i_{{nm}} {\mathcal R}^j_{mn} + {\mathcal R}^j_{{nm}} {\mathcal R}^i_{mn}\bigr)}{\omega_{mn}}.
\end{equation}
Using $\omega_{mn}=(\varepsilon_m-\varepsilon_n)/\hbar$, this expression can be rewritten entirely in terms of energy denominators and is identical to the field-induced positional shift derived in Ref.~\onlinecite{GaoYangNiu2014}. The result is gauge invariant as it depends only on products of interband Berry connections. In contrast to Ref.~\onlinecite{GaoYangNiu2014}, where the positional shift arises from a field-induced correction to the Bloch basis, here it follows directly from the dynamics encoded in the semiclassical equation of motion through the Zitterbewegung velocity.

The time-independent Berry-curvature term leads to a contribution to the electron position that is linear in time, i.e. it represents drift. By contrast, the steady-state shift $\Delta r^i_{E; m}$ is independent of time, which suggests it should be viewed as a steady-state geometric dipole induced by the electric field \cite{resta-polarization-review, GaoYangNiu2014}. Semiclassical dynamics therefore describes not only drift of the electron's center of mass but also motion \textit{relative} to the center of mass. Disentangling center-of-mass and relative motion is, in general, nontrivial.

\textit{Implications for the charge current}. Having introduced the notion of a Zitterbewegung velocity, the next question is how this new term should be treated when determining expectation values of observables. In general one does not expect the oscillatory microscopic motion itself to be directly visible in steady-state experiment. The correct strategy is to integrate over phase space first, weighted by the appropriate occupation factor. For the Zitterbewegung contribution to the charge current this means integrating $- e \sum_m {\bm \zeta}_m f_m$. Summing over occupied states causes the oscillatory contributions to dephase because different interband pairs contribute with different frequencies. Since we are interested in steady-state values, we require the asymptotic behaviour $t \rightarrow \infty$. In generic cases we expect these terms to vanish from the expectation value of the current in that limit. To see explicitly when the oscillatory Zitterbewegung terms vanish, and the important special case when they do not, we examine some models.

Consider first massive Dirac fermions described by the Hamiltonian $H_0 = \alpha (\sigma_x k_y - \sigma_y k_x) + m \sigma_z$, with $\sigma_i$ the Pauli matrices, appropriate to the surface states of a topological insulator with a magnetization $m$ \cite{Culcer2012}. The eigenvalues are $\varepsilon_\pm = \pm \sqrt{\alpha^2 k^2 + m^2}$. Consider first the chemical potential in the gap, so that only the valence band is full, and focus on the $\cos$ term in ${\bm \zeta}$. This vanishes for longitudinal transport $i = x$, while for Hall transport $i = y$ it takes the form $c_{21}^y = ieE_x ({\mathcal R}^y_{{12}} {\mathcal R}^x_{21} - {\mathcal R}^x_{{12}} {\mathcal R}^y_{21})$. Integrated over the full valence band, the asymptotic long-time limit reduces to $\sin(2mt)/(mt)$, which tends to zero in the strict steady-state limit $t \rightarrow \infty$. On the other hand, the $\sin$ term in ${\bm \zeta}$ vanishes for Hall transport. It survives for longitudinal transport, $s_{21}^x = 2eE_x |{\mathcal R}^x_{{12}} {\mathcal R}^x_{21}|$, where it integrates to $\cos(2mt)/(mt)$ at long times, which again vanishes in the steady state.

Next consider the Fermi energy in the conduction band. For Hall transport the $\cos$ term in the conduction band has the form $c_{12}^y = - c_{21}^y$ and the long-time trend is the same as above, with $m$ in the sine replaced by $E_F$. Hence this term also vanishes in the long-time limit. Finally, assuming $E_F \gg m$, the sinusoidal term in ${\bm \zeta}$ at long times again reduces to $\cos(2E_F t)/(2E_F t)$, which vanishes in the steady state. Thus, for two-dimensional massive Dirac fermions, Zitterbewegung affects the microscopic time-dependent velocity but does not produce a persistent linear-response current.

\textit{Massless Dirac fermions}. The massless limit requires separate treatment because the limits $m\to 0$ and $t\to\infty$ do not commute. This noncommutativity is physically important: for any finite mass the oscillatory contributions dephase, whereas exactly at $m=0$ the low-energy states near $k=0$ generate a singular long-time response. Hence we examine the Hamiltonian above directly at $m = 0$. In this case, since time reversal symmetry is preserved, there is no Hall transport, only longitudinal transport, where, as expected, the $\cos$ term vanishes. On the other hand, the expectation value of the $\sin$ term in ${\bm \zeta}$, multiplied by $-e$ for the charge, is:
\begin{equation}
\arraycolsep 0.3ex
\begin{array}{rl}
\displaystyle - e \zeta_x(t) = & \displaystyle -2e^2 E_x |{\mathcal R}^{12}_x|^2 \sin (2 \alpha kt/\hbar).
\end{array}
\end{equation}
Here, as above, $1$ is the conduction band and $2$ is the valence band. In the general case this must be multiplied by $(f_1 - f_2)$ and integrated over phase space. Now consider the chemical potential $\mu = 0$, add the factor of $1/(2\pi)^2$, and integrate only over the valence band, i.e. $f_1 = 0$, $f_2 = \Theta(\alpha k)$:
\begin{equation}
\arraycolsep 0.3ex
\begin{array}{rl}
\displaystyle j^{Zb}_x = & \displaystyle 2e^2 E_x \int \frac{dk \, k}{2\pi} \int\frac{d\theta}{2\pi} |{\mathcal R}^{12}_x|^2 \sin \frac{2\alpha kt}{\hbar} \, \Theta(\alpha k).
\end{array}
\end{equation}
We recover the Berry connection from the interband velocity $v_x^{12} = i (\alpha/\hbar) \sin \theta$, leading to ${\mathcal R}^{12}_x = \sin \theta/(2k)$. Substituting this gives
\begin{equation}
\arraycolsep 0.3ex
\begin{array}{rl}
\displaystyle j^{Zb}_x = & \displaystyle \frac{e^2 E_x}{4h} \int_0^{\infty} dk \, \frac{\sin (2 \alpha kt/\hbar)}{k}.
\end{array}
\end{equation}
Although $f_1 - f_2$ vanishes at $k=0$, the result is controlled by the infrared region $k \to 0$ rather than the value of the integrand at a single point. The $1/k$ kernel arising from the interband matrix elements leads to a finite contribution after integration, so that the conductivity is determined by the small-$k$ behavior. The ultraviolet details drop out of the asymptotic limit. Using
$ \int_0^{\infty} dk \, \frac{\sin (2\alpha kt/\hbar)}{k} = \frac{\pi}{2}$,
we obtain the longitudinal conductivity at the Dirac point,
$\sigma_{xx}^{Zb} = \frac{\pi e^2}{8h}$, which is the well-known minimum-conductivity result in this formulation \cite{Ludwig1994, Sinitsyn2006_II, Tworzydlo2006, Katsnelson_Zitterbewegung_EPJB2006}. Hence the minimum conductivity of Dirac fermions stems from precisely the same interband-coherence term that, more generally, produces the steady-state position shift of Bloch electrons. In contrast, when the Fermi energy lies in the conduction band the additional Zitterbewegung terms evolve as $\cos(2E_F t)/(2E_F t)$ at long times and therefore vanish in the steady-state current.

\textit{Other models}. We have considered explicitly a two-band isotropic model, where the Zitterbewegung velocity terms at finite Fermi energy dephase because the oscillation frequencies differ for different values of $k$. In such cases the interband frequency $\omega_{mn}(\bm k)$ varies across phase space, and integration over phase space produces destructive interference in the long-time limit. For more realistic band structures this dephasing is generically even stronger. Anisotropy alone ensures that $\omega_{mn}$ varies with the angular coordinate as well as the magnitude of ${\bm k}$, so angular integration already suppresses the oscillatory contribution. Effects such as warping, tilting, and particle-hole asymmetry enhance this cancellation. Furthermore, in multiband systems the number of distinct interband transition frequencies increases, strengthening the dephasing mechanism. Apart from special situations such as the massless Dirac point, the Zitterbewegung velocity therefore does not produce a persistent contribution to the steady-state linear current.

\textit{Discussion}. The central novelty of the present work is the identification of an explicit Zitterbewegung contribution $\bm\zeta_m(t)$ to the semiclassical equations of motion. This term emerges naturally when the semiclassical dynamics is derived from the time-dependent density matrix rather than from a strictly adiabatic single-band wave-packet construction \cite{GaneshSundaram1999,XiaoDi2010,Interband-Coherence-PRB-2017-Dimi}. Since the density-matrix formulation automatically incorporates multi-band effects \cite{Luttinger_AHE_PR58,Vasko2005,Interband-Coherence-PRB-2017-Dimi,JE-PRR-Rhonald-2022}, the method proposed in this work provides a powerful alternative route to deriving semiclassical equations in multi-band systems. The conventional wave-packet derivation of semiclassical dynamics is formulated in terms of a single isolated band and therefore naturally reproduces the Berry-curvature anomalous velocity, which can be written in an effective single-band form \cite{GaneshSundaram1999,XiaoDi2010}. In contrast, the Zitterbewegung velocity arises from interband coherence and cannot in general be reduced to a purely single-band quantity. For this reason, in the wave-packet approach the quantum-metric-related position shift must be evaluated separately by incorporating an adiabatic correction to the basis functions \cite{GaoYangNiu2014}. Whereas one may attempt a multi-band wave-packet formulation along the lines of Ref.~\onlinecite{Culcer-2005-WPK}, the possibility of the wave-packet splitting when the energy differences become large introduces strong caveats to such an approach. 


The velocity contains two qualitatively distinct contributions: a time-independent Berry-curvature term and an oscillatory interband term. When integrated over time, these two contributions exchange roles. The constant Berry-curvature velocity produces drift linear in time, while the oscillatory Zitterbewegung velocity produces a finite, time-independent correction to the position denoted by $\Delta r^i_{E;m}$. The apparently paradoxical feature that an oscillatory velocity can generate a finite displacement is resolved by examining the time integral of $\sin(\omega_{mn} t)$. After integration over phase space, the oscillatory $\cos(\omega_{mn} t)$ term dephases in the long-time limit, while the constant $1/\omega_{mn}$ term survives. The Zitterbewegung velocity involves both the symmetric and antisymmetric parts of the quantum geometric tensor, weighted by oscillatory factors determined by interband energy differences. In the velocity itself these contributions are transient and generally dephase after phase-space averaging. In the time-integrated dynamics, however, the oscillatory sine term produces a finite geometric displacement. Zitterbewegung therefore survives coarse graining as the dynamical origin of geometric corrections to semiclassical transport. Hence this perspective also clarifies the role of quantum geometry. 

The present approach assumes nondegenerate bands and a disorder-free system, consistent with the usual regime of validity of semiclassical transport and Kubo linear response \cite{XiaoDi2010}. The key distinction from the conventional adiabatic prescription is that the electric field is switched on at a finite time rather than in the infinitely distant past. Retaining the explicit time dependence preserves the interband dynamics responsible for the Zitterbewegung velocity and the associated position shift. This time dependence must be retained precisely in order to preserve the semiclassical point of view, in which electron dynamics is characterized in terms of position and velocity and one must be able to integrate the velocity to obtain the position. In contrast, in fully quantum-mechanical response theory, position and velocity are distinct operators and no a priori relationship between their expectation values is assumed.

\textit{Outlook}. We have demonstrated that Zitterbewegung leads directly to a new term in the semiclassical equations of motion, the \textit{Zitterbewegung velocity}. This contribution underlies the field-induced position shift and, in special cases such as massless Dirac fermions, produces observable transport effects including the minimum conductivity. The extension of this framework to disordered systems, where additional extrinsic velocity corrections arise \cite{JE-PRR-Rhonald-2022}, remains an important direction for future work.

\acknowledgments This work is supported by the Australian Research Council Discovery Project DP2401062. I am grateful to Qian Niu, Yang Gao, Di Xiao, Roberto Raimondi, Asle Sudb{\o}, Rhonald Burgos, James Cullen and Grzegorz Brz{\k e}czyszczykiewicz for enlightening discussions.

\bibliography{Zitterbewegung_extended}

\begin{thebibliography}{76}%
\makeatletter
\providecommand \@ifxundefined [1]{%
 \@ifx{#1\undefined}
}%
\providecommand \@ifnum [1]{%
 \ifnum #1\expandafter \@firstoftwo
 \else \expandafter \@secondoftwo
 \fi
}%
\providecommand \@ifx [1]{%
 \ifx #1\expandafter \@firstoftwo
 \else \expandafter \@secondoftwo
 \fi
}%
\providecommand \natexlab [1]{#1}%
\providecommand \enquote  [1]{``#1''}%
\providecommand \bibnamefont  [1]{#1}%
\providecommand \bibfnamefont [1]{#1}%
\providecommand \citenamefont [1]{#1}%
\providecommand \href@noop [0]{\@secondoftwo}%
\providecommand \href [0]{\begingroup \@sanitize@url \@href}%
\providecommand \@href[1]{\@@startlink{#1}\@@href}%
\providecommand \@@href[1]{\endgroup#1\@@endlink}%
\providecommand \@sanitize@url [0]{\catcode `\\12\catcode `\$12\catcode
  `\&12\catcode `\#12\catcode `\^12\catcode `\_12\catcode `\%12\relax}%
\providecommand \@@startlink[1]{}%
\providecommand \@@endlink[0]{}%
\providecommand \url  [0]{\begingroup\@sanitize@url \@url }%
\providecommand \@url [1]{\endgroup\@href {#1}{\urlprefix }}%
\providecommand \urlprefix  [0]{URL }%
\providecommand \Eprint [0]{\href }%
\providecommand \doibase [0]{http://dx.doi.org/}%
\providecommand \selectlanguage [0]{\@gobble}%
\providecommand \bibinfo  [0]{\@secondoftwo}%
\providecommand \bibfield  [0]{\@secondoftwo}%
\providecommand \translation [1]{[#1]}%
\providecommand \BibitemOpen [0]{}%
\providecommand \bibitemStop [0]{}%
\providecommand \bibitemNoStop [0]{.\EOS\space}%
\providecommand \EOS [0]{\spacefactor3000\relax}%
\providecommand \BibitemShut  [1]{\csname bibitem#1\endcsname}%
\let\auto@bib@innerbib\@empty
\bibitem [{\citenamefont {Chambers}(1966)}]{Chambers1966}%
  \BibitemOpen
  \bibfield  {author} {\bibinfo {author} {\bibfnamefont {R.~G.}\ \bibnamefont
  {Chambers}},\ }\href {\doibase 10.1088/0370-1328/89/4/302} {\bibfield
  {journal} {\bibinfo  {journal} {Proceedings of the Physical Society}\
  }\textbf {\bibinfo {volume} {89}},\ \bibinfo {pages} {695} (\bibinfo {year}
  {1966})}\BibitemShut {NoStop}%
\bibitem [{\citenamefont {Ashcroft}\ and\ \citenamefont
  {Mermin}(1976)}]{Ashcroft76}%
  \BibitemOpen
  \bibfield  {author} {\bibinfo {author} {\bibfnamefont {N.~W.}\ \bibnamefont
  {Ashcroft}}\ and\ \bibinfo {author} {\bibfnamefont {N.~D.}\ \bibnamefont
  {Mermin}},\ }\href@noop {} {\emph {\bibinfo {title} {{S}olid {S}tate
  {P}hysics}}}\ (\bibinfo  {publisher} {Holt-Saunders},\ \bibinfo {year}
  {1976})\BibitemShut {NoStop}%
\bibitem [{\citenamefont {Sundaram}\ and\ \citenamefont
  {Niu}(1999)}]{GaneshSundaram1999}%
  \BibitemOpen
  \bibfield  {author} {\bibinfo {author} {\bibfnamefont {G.}~\bibnamefont
  {Sundaram}}\ and\ \bibinfo {author} {\bibfnamefont {Q.}~\bibnamefont {Niu}},\
  }\href {\doibase 10.1103/PhysRevB.59.14915} {\bibfield  {journal} {\bibinfo
  {journal} {Phys. Rev. B}\ }\textbf {\bibinfo {volume} {59}},\ \bibinfo
  {pages} {14915} (\bibinfo {year} {1999})}\BibitemShut {NoStop}%
\bibitem [{\citenamefont {Jungwirth}\ \emph {et~al.}(2002)\citenamefont
  {Jungwirth}, \citenamefont {Niu},\ and\ \citenamefont
  {MacDonald}}]{Jungwirth2002}%
  \BibitemOpen
  \bibfield  {author} {\bibinfo {author} {\bibfnamefont {T.}~\bibnamefont
  {Jungwirth}}, \bibinfo {author} {\bibfnamefont {Q.}~\bibnamefont {Niu}}, \
  and\ \bibinfo {author} {\bibfnamefont {A.~H.}\ \bibnamefont {MacDonald}},\
  }\href {\doibase 10.1103/PhysRevLett.88.207208} {\bibfield  {journal}
  {\bibinfo  {journal} {Phys. Rev. Lett.}\ }\textbf {\bibinfo {volume} {88}},\
  \bibinfo {pages} {207208} (\bibinfo {year} {2002})}\BibitemShut {NoStop}%
\bibitem [{\citenamefont {Sinitsyn}(2007)}]{Sinitsyn_2007JPCM}%
  \BibitemOpen
  \bibfield  {author} {\bibinfo {author} {\bibfnamefont {N.~A.}\ \bibnamefont
  {Sinitsyn}},\ }\href {\doibase 10.1088/0953-8984/20/02/023201} {\bibfield
  {journal} {\bibinfo  {journal} {Journal of Physics: Condensed Matter}\
  }\textbf {\bibinfo {volume} {20}},\ \bibinfo {pages} {023201} (\bibinfo
  {year} {2007})}\BibitemShut {NoStop}%
\bibitem [{\citenamefont {Nagaosa}\ \emph {et~al.}(2010)\citenamefont
  {Nagaosa}, \citenamefont {Sinova}, \citenamefont {Onoda}, \citenamefont
  {MacDonald},\ and\ \citenamefont {Ong}}]{AHE-RevModPhys.82.1539}%
  \BibitemOpen
  \bibfield  {author} {\bibinfo {author} {\bibfnamefont {N.}~\bibnamefont
  {Nagaosa}}, \bibinfo {author} {\bibfnamefont {J.}~\bibnamefont {Sinova}},
  \bibinfo {author} {\bibfnamefont {S.}~\bibnamefont {Onoda}}, \bibinfo
  {author} {\bibfnamefont {A.~H.}\ \bibnamefont {MacDonald}}, \ and\ \bibinfo
  {author} {\bibfnamefont {N.~P.}\ \bibnamefont {Ong}},\ }\href {\doibase
  10.1103/RevModPhys.82.1539} {\bibfield  {journal} {\bibinfo  {journal} {Rev.
  Mod. Phys.}\ }\textbf {\bibinfo {volume} {82}},\ \bibinfo {pages} {1539}
  (\bibinfo {year} {2010})}\BibitemShut {NoStop}%
\bibitem [{\citenamefont {Matsumoto}\ and\ \citenamefont
  {Murakami}(2011)}]{Matsumoto2011}%
  \BibitemOpen
  \bibfield  {author} {\bibinfo {author} {\bibfnamefont {R.}~\bibnamefont
  {Matsumoto}}\ and\ \bibinfo {author} {\bibfnamefont {S.}~\bibnamefont
  {Murakami}},\ }\href {\doibase 10.1103/PhysRevLett.106.197202} {\bibfield
  {journal} {\bibinfo  {journal} {Phys. Rev. Lett.}\ }\textbf {\bibinfo
  {volume} {106}},\ \bibinfo {pages} {197202} (\bibinfo {year}
  {2011})}\BibitemShut {NoStop}%
\bibitem [{\citenamefont {Cheng}\ \emph {et~al.}(2016)\citenamefont {Cheng},
  \citenamefont {Okamoto},\ and\ \citenamefont {Xiao}}]{ChengRan2016}%
  \BibitemOpen
  \bibfield  {author} {\bibinfo {author} {\bibfnamefont {R.}~\bibnamefont
  {Cheng}}, \bibinfo {author} {\bibfnamefont {S.}~\bibnamefont {Okamoto}}, \
  and\ \bibinfo {author} {\bibfnamefont {D.}~\bibnamefont {Xiao}},\ }\href
  {\doibase 10.1103/PhysRevLett.117.217202} {\bibfield  {journal} {\bibinfo
  {journal} {Phys. Rev. Lett.}\ }\textbf {\bibinfo {volume} {117}},\ \bibinfo
  {pages} {217202} (\bibinfo {year} {2016})}\BibitemShut {NoStop}%
\bibitem [{\citenamefont {Gao}\ and\ \citenamefont {Xiao}(2019)}]{GaoYang2019}%
  \BibitemOpen
  \bibfield  {author} {\bibinfo {author} {\bibfnamefont {Y.}~\bibnamefont
  {Gao}}\ and\ \bibinfo {author} {\bibfnamefont {D.}~\bibnamefont {Xiao}},\
  }\href {\doibase 10.1103/PhysRevLett.122.227402} {\bibfield  {journal}
  {\bibinfo  {journal} {Phys. Rev. Lett.}\ }\textbf {\bibinfo {volume} {122}},\
  \bibinfo {pages} {227402} (\bibinfo {year} {2019})}\BibitemShut {NoStop}%
\bibitem [{\citenamefont {Zhang}\ \emph {et~al.}(2019)\citenamefont {Zhang},
  \citenamefont {Zhang}, \citenamefont {Okamoto},\ and\ \citenamefont
  {Xiao}}]{Zhang2019}%
  \BibitemOpen
  \bibfield  {author} {\bibinfo {author} {\bibfnamefont {X.}~\bibnamefont
  {Zhang}}, \bibinfo {author} {\bibfnamefont {Y.}~\bibnamefont {Zhang}},
  \bibinfo {author} {\bibfnamefont {S.}~\bibnamefont {Okamoto}}, \ and\
  \bibinfo {author} {\bibfnamefont {D.}~\bibnamefont {Xiao}},\ }\href {\doibase
  10.1103/PhysRevLett.123.167202} {\bibfield  {journal} {\bibinfo  {journal}
  {Phys. Rev. Lett.}\ }\textbf {\bibinfo {volume} {123}},\ \bibinfo {pages}
  {167202} (\bibinfo {year} {2019})}\BibitemShut {NoStop}%
\bibitem [{\citenamefont {Culcer}(2024)}]{Culcer_AHE_2024}%
  \BibitemOpen
  \bibfield  {author} {\bibinfo {author} {\bibfnamefont {D.}~\bibnamefont
  {Culcer}},\ }in\ \href {\doibase 10.1016/B978-0-323-90800-9.00006-8} {\emph
  {\bibinfo {booktitle} {Encyclopedia of Condensed Matter Physics}}},\
  Vol.~\bibinfo {volume} {1}\ (\bibinfo  {publisher} {Elsevier},\ \bibinfo
  {year} {2024})\ p.\ \bibinfo {pages} {587}\BibitemShut {NoStop}%
\bibitem [{\citenamefont {Shindou}\ and\ \citenamefont
  {Imura}(2005)}]{SHINDOU2005}%
  \BibitemOpen
  \bibfield  {author} {\bibinfo {author} {\bibfnamefont {R.}~\bibnamefont
  {Shindou}}\ and\ \bibinfo {author} {\bibfnamefont {K.-I.}\ \bibnamefont
  {Imura}},\ }\href {\doibase https://doi.org/10.1016/j.nuclphysb.2005.05.019}
  {\bibfield  {journal} {\bibinfo  {journal} {Nuclear Physics B}\ }\textbf
  {\bibinfo {volume} {720}},\ \bibinfo {pages} {399} (\bibinfo {year}
  {2005})}\BibitemShut {NoStop}%
\bibitem [{\citenamefont {Dumitrescu}\ \emph {et~al.}(2012)\citenamefont
  {Dumitrescu}, \citenamefont {Zhang}, \citenamefont {Marinescu},\ and\
  \citenamefont {Tewari}}]{Dumitrescu2012}%
  \BibitemOpen
  \bibfield  {author} {\bibinfo {author} {\bibfnamefont {E.}~\bibnamefont
  {Dumitrescu}}, \bibinfo {author} {\bibfnamefont {C.}~\bibnamefont {Zhang}},
  \bibinfo {author} {\bibfnamefont {D.~C.}\ \bibnamefont {Marinescu}}, \ and\
  \bibinfo {author} {\bibfnamefont {S.}~\bibnamefont {Tewari}},\ }\href
  {\doibase 10.1103/PhysRevB.85.245301} {\bibfield  {journal} {\bibinfo
  {journal} {Phys. Rev. B}\ }\textbf {\bibinfo {volume} {85}},\ \bibinfo
  {pages} {245301} (\bibinfo {year} {2012})}\BibitemShut {NoStop}%
\bibitem [{\citenamefont {Son}\ and\ \citenamefont {Spivak}(2013)}]{Son2013}%
  \BibitemOpen
  \bibfield  {author} {\bibinfo {author} {\bibfnamefont {D.~T.}\ \bibnamefont
  {Son}}\ and\ \bibinfo {author} {\bibfnamefont {B.~Z.}\ \bibnamefont
  {Spivak}},\ }\href {\doibase 10.1103/PhysRevB.88.104412} {\bibfield
  {journal} {\bibinfo  {journal} {Phys. Rev. B}\ }\textbf {\bibinfo {volume}
  {88}},\ \bibinfo {pages} {104412} (\bibinfo {year} {2013})}\BibitemShut
  {NoStop}%
\bibitem [{\citenamefont {Song}\ \emph {et~al.}(2015)\citenamefont {Song},
  \citenamefont {Refael},\ and\ \citenamefont {Lee}}]{SongJustin2015}%
  \BibitemOpen
  \bibfield  {author} {\bibinfo {author} {\bibfnamefont {J.~C.~W.}\
  \bibnamefont {Song}}, \bibinfo {author} {\bibfnamefont {G.}~\bibnamefont
  {Refael}}, \ and\ \bibinfo {author} {\bibfnamefont {P.~A.}\ \bibnamefont
  {Lee}},\ }\href {\doibase 10.1103/PhysRevB.92.180204} {\bibfield  {journal}
  {\bibinfo  {journal} {Phys. Rev. B}\ }\textbf {\bibinfo {volume} {92}},\
  \bibinfo {pages} {180204} (\bibinfo {year} {2015})}\BibitemShut {NoStop}%
\bibitem [{\citenamefont {Yang}\ \emph {et~al.}(2015)\citenamefont {Yang},
  \citenamefont {Pan},\ and\ \citenamefont {Zhang}}]{YangShengyuan2015}%
  \BibitemOpen
  \bibfield  {author} {\bibinfo {author} {\bibfnamefont {S.~A.}\ \bibnamefont
  {Yang}}, \bibinfo {author} {\bibfnamefont {H.}~\bibnamefont {Pan}}, \ and\
  \bibinfo {author} {\bibfnamefont {F.}~\bibnamefont {Zhang}},\ }\href
  {\doibase 10.1103/PhysRevLett.115.156603} {\bibfield  {journal} {\bibinfo
  {journal} {Phys. Rev. Lett.}\ }\textbf {\bibinfo {volume} {115}},\ \bibinfo
  {pages} {156603} (\bibinfo {year} {2015})}\BibitemShut {NoStop}%
\bibitem [{\citenamefont {Lensky}\ \emph {et~al.}(2015)\citenamefont {Lensky},
  \citenamefont {Song}, \citenamefont {Samutpraphoot},\ and\ \citenamefont
  {Levitov}}]{Lensky2015}%
  \BibitemOpen
  \bibfield  {author} {\bibinfo {author} {\bibfnamefont {Y.~D.}\ \bibnamefont
  {Lensky}}, \bibinfo {author} {\bibfnamefont {J.~C.~W.}\ \bibnamefont {Song}},
  \bibinfo {author} {\bibfnamefont {P.}~\bibnamefont {Samutpraphoot}}, \ and\
  \bibinfo {author} {\bibfnamefont {L.~S.}\ \bibnamefont {Levitov}},\ }\href
  {\doibase 10.1103/PhysRevLett.114.256601} {\bibfield  {journal} {\bibinfo
  {journal} {Phys. Rev. Lett.}\ }\textbf {\bibinfo {volume} {114}},\ \bibinfo
  {pages} {256601} (\bibinfo {year} {2015})}\BibitemShut {NoStop}%
\bibitem [{\citenamefont {Chang}\ and\ \citenamefont
  {Yang}(2015)}]{ChangMingChe2015}%
  \BibitemOpen
  \bibfield  {author} {\bibinfo {author} {\bibfnamefont {M.-C.}\ \bibnamefont
  {Chang}}\ and\ \bibinfo {author} {\bibfnamefont {M.-F.}\ \bibnamefont
  {Yang}},\ }\href {\doibase 10.1103/PhysRevB.91.115203} {\bibfield  {journal}
  {\bibinfo  {journal} {Phys. Rev. B}\ }\textbf {\bibinfo {volume} {91}},\
  \bibinfo {pages} {115203} (\bibinfo {year} {2015})}\BibitemShut {NoStop}%
\bibitem [{\citenamefont {Gorbar}\ \emph {et~al.}(2018)\citenamefont {Gorbar},
  \citenamefont {Miransky}, \citenamefont {Shovkovy},\ and\ \citenamefont
  {Sukhachov}}]{Gorbar2018}%
  \BibitemOpen
  \bibfield  {author} {\bibinfo {author} {\bibfnamefont {E.~V.}\ \bibnamefont
  {Gorbar}}, \bibinfo {author} {\bibfnamefont {V.~A.}\ \bibnamefont
  {Miransky}}, \bibinfo {author} {\bibfnamefont {I.~A.}\ \bibnamefont
  {Shovkovy}}, \ and\ \bibinfo {author} {\bibfnamefont {P.~O.}\ \bibnamefont
  {Sukhachov}},\ }\href {\doibase 10.1103/PhysRevB.98.045203} {\bibfield
  {journal} {\bibinfo  {journal} {Phys. Rev. B}\ }\textbf {\bibinfo {volume}
  {98}},\ \bibinfo {pages} {045203} (\bibinfo {year} {2018})}\BibitemShut
  {NoStop}%
\bibitem [{\citenamefont {Misaki}\ \emph {et~al.}(2018)\citenamefont {Misaki},
  \citenamefont {Miyashita},\ and\ \citenamefont {Nagaosa}}]{Misaki2018}%
  \BibitemOpen
  \bibfield  {author} {\bibinfo {author} {\bibfnamefont {K.}~\bibnamefont
  {Misaki}}, \bibinfo {author} {\bibfnamefont {S.}~\bibnamefont {Miyashita}}, \
  and\ \bibinfo {author} {\bibfnamefont {N.}~\bibnamefont {Nagaosa}},\ }\href
  {\doibase 10.1103/PhysRevB.97.075122} {\bibfield  {journal} {\bibinfo
  {journal} {Phys. Rev. B}\ }\textbf {\bibinfo {volume} {97}},\ \bibinfo
  {pages} {075122} (\bibinfo {year} {2018})}\BibitemShut {NoStop}%
\bibitem [{\citenamefont {Araki}(2018)}]{Araki2018}%
  \BibitemOpen
  \bibfield  {author} {\bibinfo {author} {\bibfnamefont {Y.}~\bibnamefont
  {Araki}},\ }\href {\doibase 10.1038/s41598-018-33655-w} {\bibfield  {journal}
  {\bibinfo  {journal} {Scientific Reports}\ }\textbf {\bibinfo {volume} {8}},\
  \bibinfo {pages} {15236} (\bibinfo {year} {2018})}\BibitemShut {NoStop}%
\bibitem [{\citenamefont {Alexandradinata}\ and\ \citenamefont
  {Glazman}(2018)}]{Alexandradinata2018}%
  \BibitemOpen
  \bibfield  {author} {\bibinfo {author} {\bibfnamefont {A.}~\bibnamefont
  {Alexandradinata}}\ and\ \bibinfo {author} {\bibfnamefont {L.}~\bibnamefont
  {Glazman}},\ }\href {\doibase 10.1103/PhysRevB.97.144422} {\bibfield
  {journal} {\bibinfo  {journal} {Phys. Rev. B}\ }\textbf {\bibinfo {volume}
  {97}},\ \bibinfo {pages} {144422} (\bibinfo {year} {2018})}\BibitemShut
  {NoStop}%
\bibitem [{\citenamefont {Das}\ and\ \citenamefont
  {Agarwal}(2019)}]{DasKamal2019}%
  \BibitemOpen
  \bibfield  {author} {\bibinfo {author} {\bibfnamefont {K.}~\bibnamefont
  {Das}}\ and\ \bibinfo {author} {\bibfnamefont {A.}~\bibnamefont {Agarwal}},\
  }\href {\doibase 10.1103/PhysRevB.99.085405} {\bibfield  {journal} {\bibinfo
  {journal} {Phys. Rev. B}\ }\textbf {\bibinfo {volume} {99}},\ \bibinfo
  {pages} {085405} (\bibinfo {year} {2019})}\BibitemShut {NoStop}%
\bibitem [{\citenamefont {Nandy}\ \emph {et~al.}(2019)\citenamefont {Nandy},
  \citenamefont {Taraphder},\ and\ \citenamefont {Tewari}}]{Nandy2019}%
  \BibitemOpen
  \bibfield  {author} {\bibinfo {author} {\bibfnamefont {S.}~\bibnamefont
  {Nandy}}, \bibinfo {author} {\bibfnamefont {A.}~\bibnamefont {Taraphder}}, \
  and\ \bibinfo {author} {\bibfnamefont {S.}~\bibnamefont {Tewari}},\ }\href
  {\doibase 10.1103/PhysRevB.100.115139} {\bibfield  {journal} {\bibinfo
  {journal} {Phys. Rev. B}\ }\textbf {\bibinfo {volume} {100}},\ \bibinfo
  {pages} {115139} (\bibinfo {year} {2019})}\BibitemShut {NoStop}%
\bibitem [{\citenamefont {Hou}\ \emph {et~al.}(2019)\citenamefont {Hou},
  \citenamefont {Zhou}, \citenamefont {Yang},\ and\ \citenamefont
  {Sun}}]{Hou2019}%
  \BibitemOpen
  \bibfield  {author} {\bibinfo {author} {\bibfnamefont {Z.}~\bibnamefont
  {Hou}}, \bibinfo {author} {\bibfnamefont {Y.-F.}\ \bibnamefont {Zhou}},
  \bibinfo {author} {\bibfnamefont {N.-X.}\ \bibnamefont {Yang}}, \ and\
  \bibinfo {author} {\bibfnamefont {Q.-F.}\ \bibnamefont {Sun}},\ }\href
  {\doibase 10.1038/s42005-019-0186-9} {\bibfield  {journal} {\bibinfo
  {journal} {Communications Physics}\ }\textbf {\bibinfo {volume} {2}},\
  \bibinfo {pages} {86} (\bibinfo {year} {2019})}\BibitemShut {NoStop}%
\bibitem [{\citenamefont {Zhu}\ \emph {et~al.}(2021)\citenamefont {Zhu},
  \citenamefont {Hughes},\ and\ \citenamefont
  {Alexandradinata}}]{ZhuPenghao2021}%
  \BibitemOpen
  \bibfield  {author} {\bibinfo {author} {\bibfnamefont {P.}~\bibnamefont
  {Zhu}}, \bibinfo {author} {\bibfnamefont {T.~L.}\ \bibnamefont {Hughes}}, \
  and\ \bibinfo {author} {\bibfnamefont {A.}~\bibnamefont {Alexandradinata}},\
  }\href {\doibase 10.1103/PhysRevB.103.014417} {\bibfield  {journal} {\bibinfo
   {journal} {Phys. Rev. B}\ }\textbf {\bibinfo {volume} {103}},\ \bibinfo
  {pages} {014417} (\bibinfo {year} {2021})}\BibitemShut {NoStop}%
\bibitem [{\citenamefont {Yokoyama}(2021)}]{Yokoyama2021}%
  \BibitemOpen
  \bibfield  {author} {\bibinfo {author} {\bibfnamefont {T.}~\bibnamefont
  {Yokoyama}},\ }\href {\doibase 10.1038/s41598-021-91436-4} {\bibfield
  {journal} {\bibinfo  {journal} {Sci Rep}\ }\textbf {\bibinfo {volume} {11}},\
  \bibinfo {pages} {12065} (\bibinfo {year} {2021})}\BibitemShut {NoStop}%
\bibitem [{\citenamefont {Wang}\ \emph {et~al.}(2021)\citenamefont {Wang},
  \citenamefont {Dong}, \citenamefont {Xiao},\ and\ \citenamefont
  {Niu}}]{WangZhi2021}%
  \BibitemOpen
  \bibfield  {author} {\bibinfo {author} {\bibfnamefont {Z.}~\bibnamefont
  {Wang}}, \bibinfo {author} {\bibfnamefont {L.}~\bibnamefont {Dong}}, \bibinfo
  {author} {\bibfnamefont {C.}~\bibnamefont {Xiao}}, \ and\ \bibinfo {author}
  {\bibfnamefont {Q.}~\bibnamefont {Niu}},\ }\href {\doibase
  10.1103/PhysRevLett.126.187001} {\bibfield  {journal} {\bibinfo  {journal}
  {Phys. Rev. Lett.}\ }\textbf {\bibinfo {volume} {126}},\ \bibinfo {pages}
  {187001} (\bibinfo {year} {2021})}\BibitemShut {NoStop}%
\bibitem [{\citenamefont {Das}\ and\ \citenamefont
  {Agarwal}(2021)}]{DasKamal2021}%
  \BibitemOpen
  \bibfield  {author} {\bibinfo {author} {\bibfnamefont {K.}~\bibnamefont
  {Das}}\ and\ \bibinfo {author} {\bibfnamefont {A.}~\bibnamefont {Agarwal}},\
  }\href {\doibase 10.1103/PhysRevB.103.125432} {\bibfield  {journal} {\bibinfo
   {journal} {Phys. Rev. B}\ }\textbf {\bibinfo {volume} {103}},\ \bibinfo
  {pages} {125432} (\bibinfo {year} {2021})}\BibitemShut {NoStop}%
\bibitem [{\citenamefont {Chaudhary}\ \emph {et~al.}(2021)\citenamefont
  {Chaudhary}, \citenamefont {Knapp},\ and\ \citenamefont
  {Refael}}]{Chaudhary2021}%
  \BibitemOpen
  \bibfield  {author} {\bibinfo {author} {\bibfnamefont {S.}~\bibnamefont
  {Chaudhary}}, \bibinfo {author} {\bibfnamefont {C.}~\bibnamefont {Knapp}}, \
  and\ \bibinfo {author} {\bibfnamefont {G.}~\bibnamefont {Refael}},\ }\href
  {\doibase 10.1103/PhysRevB.103.165119} {\bibfield  {journal} {\bibinfo
  {journal} {Phys. Rev. B}\ }\textbf {\bibinfo {volume} {103}},\ \bibinfo
  {pages} {165119} (\bibinfo {year} {2021})}\BibitemShut {NoStop}%
\bibitem [{\citenamefont {Bhowal}\ and\ \citenamefont
  {Vignale}(2021)}]{Bhowal2021}%
  \BibitemOpen
  \bibfield  {author} {\bibinfo {author} {\bibfnamefont {S.}~\bibnamefont
  {Bhowal}}\ and\ \bibinfo {author} {\bibfnamefont {G.}~\bibnamefont
  {Vignale}},\ }\href {\doibase 10.1103/PhysRevB.103.195309} {\bibfield
  {journal} {\bibinfo  {journal} {Phys. Rev. B}\ }\textbf {\bibinfo {volume}
  {103}},\ \bibinfo {pages} {195309} (\bibinfo {year} {2021})}\BibitemShut
  {NoStop}%
\bibitem [{\citenamefont {Battiato}(2021)}]{Battiato2021}%
  \BibitemOpen
  \bibfield  {author} {\bibinfo {author} {\bibfnamefont {M.}~\bibnamefont
  {Battiato}},\ }\href {\doibase 10.1088/1361-6404/abe2d3} {\bibfield
  {journal} {\bibinfo  {journal} {European Journal of Physics}\ }\textbf
  {\bibinfo {volume} {42}},\ \bibinfo {pages} {035403} (\bibinfo {year}
  {2021})}\BibitemShut {NoStop}%
\bibitem [{\citenamefont {Valet}\ and\ \citenamefont
  {Raimondi}(2023)}]{Raimondi_EPL2023}%
  \BibitemOpen
  \bibfield  {author} {\bibinfo {author} {\bibfnamefont {T.}~\bibnamefont
  {Valet}}\ and\ \bibinfo {author} {\bibfnamefont {R.}~\bibnamefont
  {Raimondi}},\ }\href {\doibase 10.1209/0295-5075/ace379} {\bibfield
  {journal} {\bibinfo  {journal} {Europhysics Letters (EPL)}\ }\textbf
  {\bibinfo {volume} {143}},\ \bibinfo {pages} {26004} (\bibinfo {year}
  {2023})}\BibitemShut {NoStop}%
\bibitem [{\citenamefont {Moore}\ and\ \citenamefont
  {Orenstein}(2010)}]{Moore2010}%
  \BibitemOpen
  \bibfield  {author} {\bibinfo {author} {\bibfnamefont {J.~E.}\ \bibnamefont
  {Moore}}\ and\ \bibinfo {author} {\bibfnamefont {J.}~\bibnamefont
  {Orenstein}},\ }\href {\doibase 10.1103/PhysRevLett.105.026805} {\bibfield
  {journal} {\bibinfo  {journal} {Phys. Rev. Lett.}\ }\textbf {\bibinfo
  {volume} {105}},\ \bibinfo {pages} {026805} (\bibinfo {year}
  {2010})}\BibitemShut {NoStop}%
\bibitem [{\citenamefont {Sodemann}\ and\ \citenamefont
  {Fu}(2015)}]{Sodemann2015}%
  \BibitemOpen
  \bibfield  {author} {\bibinfo {author} {\bibfnamefont {I.}~\bibnamefont
  {Sodemann}}\ and\ \bibinfo {author} {\bibfnamefont {L.}~\bibnamefont {Fu}},\
  }\href {\doibase 10.1103/PhysRevLett.115.216806} {\bibfield  {journal}
  {\bibinfo  {journal} {Phys. Rev. Lett.}\ }\textbf {\bibinfo {volume} {115}},\
  \bibinfo {pages} {216806} (\bibinfo {year} {2015})}\BibitemShut {NoStop}%
\bibitem [{\citenamefont {Morimoto}\ \emph {et~al.}(2016)\citenamefont
  {Morimoto}, \citenamefont {Zhong}, \citenamefont {Orenstein},\ and\
  \citenamefont {Moore}}]{Morimoto2016}%
  \BibitemOpen
  \bibfield  {author} {\bibinfo {author} {\bibfnamefont {T.}~\bibnamefont
  {Morimoto}}, \bibinfo {author} {\bibfnamefont {S.}~\bibnamefont {Zhong}},
  \bibinfo {author} {\bibfnamefont {J.}~\bibnamefont {Orenstein}}, \ and\
  \bibinfo {author} {\bibfnamefont {J.~E.}\ \bibnamefont {Moore}},\ }\href
  {\doibase 10.1103/PhysRevB.94.245121} {\bibfield  {journal} {\bibinfo
  {journal} {Phys. Rev. B}\ }\textbf {\bibinfo {volume} {94}},\ \bibinfo
  {pages} {245121} (\bibinfo {year} {2016})}\BibitemShut {NoStop}%
\bibitem [{\citenamefont {Zeng}\ \emph {et~al.}(2019)\citenamefont {Zeng},
  \citenamefont {Nandy}, \citenamefont {Taraphder},\ and\ \citenamefont
  {Tewari}}]{ZhenChuanchang2019}%
  \BibitemOpen
  \bibfield  {author} {\bibinfo {author} {\bibfnamefont {C.}~\bibnamefont
  {Zeng}}, \bibinfo {author} {\bibfnamefont {S.}~\bibnamefont {Nandy}},
  \bibinfo {author} {\bibfnamefont {A.}~\bibnamefont {Taraphder}}, \ and\
  \bibinfo {author} {\bibfnamefont {S.}~\bibnamefont {Tewari}},\ }\href
  {\doibase 10.1103/PhysRevB.100.245102} {\bibfield  {journal} {\bibinfo
  {journal} {Phys. Rev. B}\ }\textbf {\bibinfo {volume} {100}},\ \bibinfo
  {pages} {245102} (\bibinfo {year} {2019})}\BibitemShut {NoStop}%
\bibitem [{\citenamefont {Golub}\ \emph {et~al.}(2020)\citenamefont {Golub},
  \citenamefont {Ivchenko},\ and\ \citenamefont {Spivak}}]{Golub2020}%
  \BibitemOpen
  \bibfield  {author} {\bibinfo {author} {\bibfnamefont {L.~E.}\ \bibnamefont
  {Golub}}, \bibinfo {author} {\bibfnamefont {E.~L.}\ \bibnamefont {Ivchenko}},
  \ and\ \bibinfo {author} {\bibfnamefont {B.}~\bibnamefont {Spivak}},\ }\href
  {\doibase 10.1103/PhysRevB.102.085202} {\bibfield  {journal} {\bibinfo
  {journal} {Phys. Rev. B}\ }\textbf {\bibinfo {volume} {102}},\ \bibinfo
  {pages} {085202} (\bibinfo {year} {2020})}\BibitemShut {NoStop}%
\bibitem [{\citenamefont {Wang}\ \emph {et~al.}(2006)\citenamefont {Wang},
  \citenamefont {Yates}, \citenamefont {Souza},\ and\ \citenamefont
  {Vanderbilt}}]{Wang2006}%
  \BibitemOpen
  \bibfield  {author} {\bibinfo {author} {\bibfnamefont {X.}~\bibnamefont
  {Wang}}, \bibinfo {author} {\bibfnamefont {J.~R.}\ \bibnamefont {Yates}},
  \bibinfo {author} {\bibfnamefont {I.}~\bibnamefont {Souza}}, \ and\ \bibinfo
  {author} {\bibfnamefont {D.}~\bibnamefont {Vanderbilt}},\ }\href {\doibase
  10.1103/PhysRevB.74.195118} {\bibfield  {journal} {\bibinfo  {journal} {Phys.
  Rev. B}\ }\textbf {\bibinfo {volume} {74}},\ \bibinfo {pages} {195118}
  (\bibinfo {year} {2006})}\BibitemShut {NoStop}%
\bibitem [{\citenamefont {Wang}\ \emph {et~al.}(2007)\citenamefont {Wang},
  \citenamefont {Vanderbilt}, \citenamefont {Yates},\ and\ \citenamefont
  {Souza}}]{Wang2007}%
  \BibitemOpen
  \bibfield  {author} {\bibinfo {author} {\bibfnamefont {X.}~\bibnamefont
  {Wang}}, \bibinfo {author} {\bibfnamefont {D.}~\bibnamefont {Vanderbilt}},
  \bibinfo {author} {\bibfnamefont {J.~R.}\ \bibnamefont {Yates}}, \ and\
  \bibinfo {author} {\bibfnamefont {I.}~\bibnamefont {Souza}},\ }\href
  {\doibase 10.1103/PhysRevB.76.195109} {\bibfield  {journal} {\bibinfo
  {journal} {Phys. Rev. B}\ }\textbf {\bibinfo {volume} {76}},\ \bibinfo
  {pages} {195109} (\bibinfo {year} {2007})}\BibitemShut {NoStop}%
\bibitem [{\citenamefont {Gradhand}\ \emph {et~al.}(2012)\citenamefont
  {Gradhand}, \citenamefont {Fedorov}, \citenamefont {Pientka}, \citenamefont
  {Zahn}, \citenamefont {Mertig},\ and\ \citenamefont
  {Gy{\"o}rffy}}]{Gradhand_2012}%
  \BibitemOpen
  \bibfield  {author} {\bibinfo {author} {\bibfnamefont {M.}~\bibnamefont
  {Gradhand}}, \bibinfo {author} {\bibfnamefont {D.~V.}\ \bibnamefont
  {Fedorov}}, \bibinfo {author} {\bibfnamefont {F.}~\bibnamefont {Pientka}},
  \bibinfo {author} {\bibfnamefont {P.}~\bibnamefont {Zahn}}, \bibinfo {author}
  {\bibfnamefont {I.}~\bibnamefont {Mertig}}, \ and\ \bibinfo {author}
  {\bibfnamefont {B.~L.}\ \bibnamefont {Gy{\"o}rffy}},\ }\href {\doibase
  10.1088/0953-8984/24/21/213202} {\bibfield  {journal} {\bibinfo  {journal}
  {Journal of Physics: Condensed Matter}\ }\textbf {\bibinfo {volume} {24}},\
  \bibinfo {pages} {213202} (\bibinfo {year} {2012})}\BibitemShut {NoStop}%
\bibitem [{\citenamefont {He}\ \emph {et~al.}(2012)\citenamefont {He},
  \citenamefont {Moore},\ and\ \citenamefont {Varma}}]{HeYan2012}%
  \BibitemOpen
  \bibfield  {author} {\bibinfo {author} {\bibfnamefont {Y.}~\bibnamefont
  {He}}, \bibinfo {author} {\bibfnamefont {J.}~\bibnamefont {Moore}}, \ and\
  \bibinfo {author} {\bibfnamefont {C.~M.}\ \bibnamefont {Varma}},\ }\href
  {\doibase 10.1103/PhysRevB.85.155106} {\bibfield  {journal} {\bibinfo
  {journal} {Phys. Rev. B}\ }\textbf {\bibinfo {volume} {85}},\ \bibinfo
  {pages} {155106} (\bibinfo {year} {2012})}\BibitemShut {NoStop}%
\bibitem [{\citenamefont {Chen}\ \emph {et~al.}(2014)\citenamefont {Chen},
  \citenamefont {Niu},\ and\ \citenamefont {MacDonald}}]{ChenHua2014}%
  \BibitemOpen
  \bibfield  {author} {\bibinfo {author} {\bibfnamefont {H.}~\bibnamefont
  {Chen}}, \bibinfo {author} {\bibfnamefont {Q.}~\bibnamefont {Niu}}, \ and\
  \bibinfo {author} {\bibfnamefont {A.~H.}\ \bibnamefont {MacDonald}},\ }\href
  {\doibase 10.1103/PhysRevLett.112.017205} {\bibfield  {journal} {\bibinfo
  {journal} {Phys. Rev. Lett.}\ }\textbf {\bibinfo {volume} {112}},\ \bibinfo
  {pages} {017205} (\bibinfo {year} {2014})}\BibitemShut {NoStop}%
\bibitem [{\citenamefont {Olsen}\ and\ \citenamefont
  {Souza}(2015)}]{Olsen2015}%
  \BibitemOpen
  \bibfield  {author} {\bibinfo {author} {\bibfnamefont {T.}~\bibnamefont
  {Olsen}}\ and\ \bibinfo {author} {\bibfnamefont {I.}~\bibnamefont {Souza}},\
  }\href {\doibase 10.1103/PhysRevB.92.125146} {\bibfield  {journal} {\bibinfo
  {journal} {Phys. Rev. B}\ }\textbf {\bibinfo {volume} {92}},\ \bibinfo
  {pages} {125146} (\bibinfo {year} {2015})}\BibitemShut {NoStop}%
\bibitem [{\citenamefont {Feng}\ \emph {et~al.}(2016)\citenamefont {Feng},
  \citenamefont {Liu}, \citenamefont {Liu}, \citenamefont {Zhou},\ and\
  \citenamefont {Yao}}]{WanxiangFeng2016}%
  \BibitemOpen
  \bibfield  {author} {\bibinfo {author} {\bibfnamefont {W.}~\bibnamefont
  {Feng}}, \bibinfo {author} {\bibfnamefont {C.-C.}\ \bibnamefont {Liu}},
  \bibinfo {author} {\bibfnamefont {G.-B.}\ \bibnamefont {Liu}}, \bibinfo
  {author} {\bibfnamefont {J.-J.}\ \bibnamefont {Zhou}}, \ and\ \bibinfo
  {author} {\bibfnamefont {Y.}~\bibnamefont {Yao}},\ }\href {\doibase
  https://doi.org/10.1016/j.commatsci.2015.09.020} {\bibfield  {journal}
  {\bibinfo  {journal} {Computational Materials Science}\ }\textbf {\bibinfo
  {volume} {112}},\ \bibinfo {pages} {428} (\bibinfo {year} {2016})},\ \bibinfo
  {note} {computational Materials Science in China}\BibitemShut {NoStop}%
\bibitem [{\citenamefont {Dai}\ \emph {et~al.}(2017)\citenamefont {Dai},
  \citenamefont {Du},\ and\ \citenamefont {Lu}}]{XinDai2017}%
  \BibitemOpen
  \bibfield  {author} {\bibinfo {author} {\bibfnamefont {X.}~\bibnamefont
  {Dai}}, \bibinfo {author} {\bibfnamefont {Z.~Z.}\ \bibnamefont {Du}}, \ and\
  \bibinfo {author} {\bibfnamefont {H.-Z.}\ \bibnamefont {Lu}},\ }\href
  {\doibase 10.1103/PhysRevLett.119.166601} {\bibfield  {journal} {\bibinfo
  {journal} {Phys. Rev. Lett.}\ }\textbf {\bibinfo {volume} {119}},\ \bibinfo
  {pages} {166601} (\bibinfo {year} {2017})}\BibitemShut {NoStop}%
\bibitem [{\citenamefont {Martiny}\ \emph {et~al.}(2019)\citenamefont
  {Martiny}, \citenamefont {Kaasbjerg},\ and\ \citenamefont
  {Jauho}}]{Martiny2019}%
  \BibitemOpen
  \bibfield  {author} {\bibinfo {author} {\bibfnamefont {J.~H.~J.}\
  \bibnamefont {Martiny}}, \bibinfo {author} {\bibfnamefont {K.}~\bibnamefont
  {Kaasbjerg}}, \ and\ \bibinfo {author} {\bibfnamefont {A.-P.}\ \bibnamefont
  {Jauho}},\ }\href {\doibase 10.1103/PhysRevB.100.155414} {\bibfield
  {journal} {\bibinfo  {journal} {Phys. Rev. B}\ }\textbf {\bibinfo {volume}
  {100}},\ \bibinfo {pages} {155414} (\bibinfo {year} {2019})}\BibitemShut
  {NoStop}%
\bibitem [{\citenamefont {Wuttke}\ \emph {et~al.}(2019)\citenamefont {Wuttke},
  \citenamefont {Caglieris}, \citenamefont {Sykora}, \citenamefont
  {Scaravaggi}, \citenamefont {Wolter}, \citenamefont {Manna}, \citenamefont
  {S\"uss}, \citenamefont {Shekhar}, \citenamefont {Felser}, \citenamefont
  {B\"uchner},\ and\ \citenamefont {Hess}}]{Wuttke2019}%
  \BibitemOpen
  \bibfield  {author} {\bibinfo {author} {\bibfnamefont {C.}~\bibnamefont
  {Wuttke}}, \bibinfo {author} {\bibfnamefont {F.}~\bibnamefont {Caglieris}},
  \bibinfo {author} {\bibfnamefont {S.}~\bibnamefont {Sykora}}, \bibinfo
  {author} {\bibfnamefont {F.}~\bibnamefont {Scaravaggi}}, \bibinfo {author}
  {\bibfnamefont {A.~U.~B.}\ \bibnamefont {Wolter}}, \bibinfo {author}
  {\bibfnamefont {K.}~\bibnamefont {Manna}}, \bibinfo {author} {\bibfnamefont
  {V.}~\bibnamefont {S\"uss}}, \bibinfo {author} {\bibfnamefont
  {C.}~\bibnamefont {Shekhar}}, \bibinfo {author} {\bibfnamefont
  {C.}~\bibnamefont {Felser}}, \bibinfo {author} {\bibfnamefont
  {B.}~\bibnamefont {B\"uchner}}, \ and\ \bibinfo {author} {\bibfnamefont
  {C.}~\bibnamefont {Hess}},\ }\href {\doibase 10.1103/PhysRevB.100.085111}
  {\bibfield  {journal} {\bibinfo  {journal} {Phys. Rev. B}\ }\textbf {\bibinfo
  {volume} {100}},\ \bibinfo {pages} {085111} (\bibinfo {year}
  {2019})}\BibitemShut {NoStop}%
\bibitem [{\citenamefont {Du}\ \emph {et~al.}(2020)\citenamefont {Du},
  \citenamefont {Tang}, \citenamefont {Li}, \citenamefont {Lin}, \citenamefont
  {Xu}, \citenamefont {Duan},\ and\ \citenamefont {Rubio}}]{DuShiqiao2020}%
  \BibitemOpen
  \bibfield  {author} {\bibinfo {author} {\bibfnamefont {S.}~\bibnamefont
  {Du}}, \bibinfo {author} {\bibfnamefont {P.}~\bibnamefont {Tang}}, \bibinfo
  {author} {\bibfnamefont {J.}~\bibnamefont {Li}}, \bibinfo {author}
  {\bibfnamefont {Z.}~\bibnamefont {Lin}}, \bibinfo {author} {\bibfnamefont
  {Y.}~\bibnamefont {Xu}}, \bibinfo {author} {\bibfnamefont {W.}~\bibnamefont
  {Duan}}, \ and\ \bibinfo {author} {\bibfnamefont {A.}~\bibnamefont {Rubio}},\
  }\href {\doibase 10.1103/PhysRevResearch.2.022025} {\bibfield  {journal}
  {\bibinfo  {journal} {Phys. Rev. Research}\ }\textbf {\bibinfo {volume}
  {2}},\ \bibinfo {pages} {022025} (\bibinfo {year} {2020})}\BibitemShut
  {NoStop}%
\bibitem [{\citenamefont {He}\ \emph {et~al.}(2020)\citenamefont {He},
  \citenamefont {Goldhaber-Gordon},\ and\ \citenamefont {Law}}]{KTLaw2020}%
  \BibitemOpen
  \bibfield  {author} {\bibinfo {author} {\bibfnamefont {W.-Y.}\ \bibnamefont
  {He}}, \bibinfo {author} {\bibfnamefont {D.}~\bibnamefont
  {Goldhaber-Gordon}}, \ and\ \bibinfo {author} {\bibfnamefont {K.~T.}\
  \bibnamefont {Law}},\ }\href {\doibase 10.1038/s41467-020-15473-9} {\bibfield
   {journal} {\bibinfo  {journal} {Nature Communications}\ }\textbf {\bibinfo
  {volume} {11}},\ \bibinfo {pages} {1650} (\bibinfo {year}
  {2020})}\BibitemShut {NoStop}%
\bibitem [{\citenamefont {He}\ and\ \citenamefont {Law}(2021)}]{HeWen2021}%
  \BibitemOpen
  \bibfield  {author} {\bibinfo {author} {\bibfnamefont {W.-Y.}\ \bibnamefont
  {He}}\ and\ \bibinfo {author} {\bibfnamefont {K.~T.}\ \bibnamefont {Law}},\
  }\href {\doibase 10.1103/PhysRevResearch.3.L032012} {\bibfield  {journal}
  {\bibinfo  {journal} {Phys. Rev. Research}\ }\textbf {\bibinfo {volume}
  {3}},\ \bibinfo {pages} {L032012} (\bibinfo {year} {2021})}\BibitemShut
  {NoStop}%
\bibitem [{\citenamefont {Chang}\ and\ \citenamefont {Niu}(2008)}]{Chang_2008}%
  \BibitemOpen
  \bibfield  {author} {\bibinfo {author} {\bibfnamefont {M.-C.}\ \bibnamefont
  {Chang}}\ and\ \bibinfo {author} {\bibfnamefont {Q.}~\bibnamefont {Niu}},\
  }\href {\doibase 10.1088/0953-8984/20/19/193202} {\bibfield  {journal}
  {\bibinfo  {journal} {Journal of Physics: Condensed Matter}\ }\textbf
  {\bibinfo {volume} {20}},\ \bibinfo {pages} {193202} (\bibinfo {year}
  {2008})}\BibitemShut {NoStop}%
\bibitem [{\citenamefont {Culcer}\ \emph {et~al.}(2017)\citenamefont {Culcer},
  \citenamefont {Sekine},\ and\ \citenamefont
  {MacDonald}}]{Interband-Coherence-PRB-2017-Dimi}%
  \BibitemOpen
  \bibfield  {author} {\bibinfo {author} {\bibfnamefont {D.}~\bibnamefont
  {Culcer}}, \bibinfo {author} {\bibfnamefont {A.}~\bibnamefont {Sekine}}, \
  and\ \bibinfo {author} {\bibfnamefont {A.~H.}\ \bibnamefont {MacDonald}},\
  }\href {\doibase 10.1103/PhysRevB.96.035106} {\bibfield  {journal} {\bibinfo
  {journal} {Phys. Rev. B}\ }\textbf {\bibinfo {volume} {96}},\ \bibinfo
  {pages} {035106} (\bibinfo {year} {2017})}\BibitemShut {NoStop}%
\bibitem [{\citenamefont {Stedman}\ \emph {et~al.}(2019)\citenamefont
  {Stedman}, \citenamefont {Timm},\ and\ \citenamefont {Woods}}]{Stedman_2019}%
  \BibitemOpen
  \bibfield  {author} {\bibinfo {author} {\bibfnamefont {T.}~\bibnamefont
  {Stedman}}, \bibinfo {author} {\bibfnamefont {C.}~\bibnamefont {Timm}}, \
  and\ \bibinfo {author} {\bibfnamefont {L.~M.}\ \bibnamefont {Woods}},\ }\href
  {\doibase 10.1088/1367-2630/ab3f5f} {\bibfield  {journal} {\bibinfo
  {journal} {New Journal of Physics}\ }\textbf {\bibinfo {volume} {21}},\
  \bibinfo {pages} {103007} (\bibinfo {year} {2019})}\BibitemShut {NoStop}%
\bibitem [{\citenamefont {Stedman}\ and\ \citenamefont
  {Woods}(2020)}]{Stedman2020}%
  \BibitemOpen
  \bibfield  {author} {\bibinfo {author} {\bibfnamefont {T.}~\bibnamefont
  {Stedman}}\ and\ \bibinfo {author} {\bibfnamefont {L.~M.}\ \bibnamefont
  {Woods}},\ }\href {\doibase 10.1103/PhysRevResearch.2.033086} {\bibfield
  {journal} {\bibinfo  {journal} {Phys. Rev. Research}\ }\textbf {\bibinfo
  {volume} {2}},\ \bibinfo {pages} {033086} (\bibinfo {year}
  {2020})}\BibitemShut {NoStop}%
\bibitem [{\citenamefont {Atencia}\ \emph {et~al.}(2022)\citenamefont
  {Atencia}, \citenamefont {Niu},\ and\ \citenamefont
  {Culcer}}]{JE-PRR-Rhonald-2022}%
  \BibitemOpen
  \bibfield  {author} {\bibinfo {author} {\bibfnamefont {R.~B.}\ \bibnamefont
  {Atencia}}, \bibinfo {author} {\bibfnamefont {Q.}~\bibnamefont {Niu}}, \ and\
  \bibinfo {author} {\bibfnamefont {D.}~\bibnamefont {Culcer}},\ }\href
  {\doibase 10.1103/PhysRevResearch.4.013001} {\bibfield  {journal} {\bibinfo
  {journal} {Phys. Rev. Res.}\ }\textbf {\bibinfo {volume} {4}},\ \bibinfo
  {pages} {013001} (\bibinfo {year} {2022})}\BibitemShut {NoStop}%
\bibitem [{\citenamefont {Gao}\ \emph {et~al.}(2014)\citenamefont {Gao},
  \citenamefont {Yang},\ and\ \citenamefont {Niu}}]{GaoYangNiu2014}%
  \BibitemOpen
  \bibfield  {author} {\bibinfo {author} {\bibfnamefont {Y.}~\bibnamefont
  {Gao}}, \bibinfo {author} {\bibfnamefont {S.~A.}\ \bibnamefont {Yang}}, \
  and\ \bibinfo {author} {\bibfnamefont {Q.}~\bibnamefont {Niu}},\ }\href
  {\doibase 10.1103/PhysRevLett.112.166601} {\bibfield  {journal} {\bibinfo
  {journal} {Phys. Rev. Lett.}\ }\textbf {\bibinfo {volume} {112}},\ \bibinfo
  {pages} {166601} (\bibinfo {year} {2014})}\BibitemShut {NoStop}%
\bibitem [{\citenamefont {Xiao}\ \emph {et~al.}(2010)\citenamefont {Xiao},
  \citenamefont {Chang},\ and\ \citenamefont {Niu}}]{XiaoDi2010}%
  \BibitemOpen
  \bibfield  {author} {\bibinfo {author} {\bibfnamefont {D.}~\bibnamefont
  {Xiao}}, \bibinfo {author} {\bibfnamefont {M.-C.}\ \bibnamefont {Chang}}, \
  and\ \bibinfo {author} {\bibfnamefont {Q.}~\bibnamefont {Niu}},\ }\href
  {\doibase 10.1103/RevModPhys.82.1959} {\bibfield  {journal} {\bibinfo
  {journal} {Rev. Mod. Phys.}\ }\textbf {\bibinfo {volume} {82}},\ \bibinfo
  {pages} {1959} (\bibinfo {year} {2010})}\BibitemShut {NoStop}%
\bibitem [{\citenamefont {Zawadzki}(2005)}]{Zawadzki_Zitterbewegung_PRB2005}%
  \BibitemOpen
  \bibfield  {author} {\bibinfo {author} {\bibfnamefont {W.}~\bibnamefont
  {Zawadzki}},\ }\href@noop {} {\bibfield  {journal} {\bibinfo  {journal}
  {Phys. Rev. B}\ }\textbf {\bibinfo {volume} {72}},\ \bibinfo {pages} {085217}
  (\bibinfo {year} {2005})}\BibitemShut {NoStop}%
\bibitem [{\citenamefont {Schliemann}\ \emph {et~al.}(2006)\citenamefont
  {Schliemann}, \citenamefont {Loss},\ and\ \citenamefont
  {Westervelt}}]{Schliemann2006}%
  \BibitemOpen
  \bibfield  {author} {\bibinfo {author} {\bibfnamefont {J.}~\bibnamefont
  {Schliemann}}, \bibinfo {author} {\bibfnamefont {D.}~\bibnamefont {Loss}}, \
  and\ \bibinfo {author} {\bibfnamefont {R.~M.}\ \bibnamefont {Westervelt}},\
  }\href@noop {} {\bibfield  {journal} {\bibinfo  {journal} {Phys. Rev. B}\
  }\textbf {\bibinfo {volume} {73}},\ \bibinfo {pages} {085323} (\bibinfo
  {year} {2006})}\BibitemShut {NoStop}%
\bibitem [{\citenamefont {Culcer}\ \emph {et~al.}(2006)\citenamefont {Culcer},
  \citenamefont {Lechner},\ and\ \citenamefont {Winkler}}]{Culcer2006}%
  \BibitemOpen
  \bibfield  {author} {\bibinfo {author} {\bibfnamefont {D.}~\bibnamefont
  {Culcer}}, \bibinfo {author} {\bibfnamefont {C.}~\bibnamefont {Lechner}}, \
  and\ \bibinfo {author} {\bibfnamefont {R.}~\bibnamefont {Winkler}},\ }\href
  {\doibase 10.1103/PhysRevLett.97.106601} {\bibfield  {journal} {\bibinfo
  {journal} {Phys. Rev. Lett.}\ }\textbf {\bibinfo {volume} {97}},\ \bibinfo
  {pages} {106601} (\bibinfo {year} {2006})}\BibitemShut {NoStop}%
\bibitem [{\citenamefont
  {Katsnelson}(2006)}]{Katsnelson_Zitterbewegung_EPJB2006}%
  \BibitemOpen
  \bibfield  {author} {\bibinfo {author} {\bibfnamefont {M.~I.}\ \bibnamefont
  {Katsnelson}},\ }\href@noop {} {\bibfield  {journal} {\bibinfo  {journal}
  {Eur. Phys. J. B}\ }\textbf {\bibinfo {volume} {51}},\ \bibinfo {pages} {157}
  (\bibinfo {year} {2006})}\BibitemShut {NoStop}%
\bibitem [{\citenamefont {Winkler}\ \emph {et~al.}(2006)\citenamefont
  {Winkler}, \citenamefont {Zulicke},\ and\ \citenamefont
  {Bolte}}]{WinklerZuelicke2007}%
  \BibitemOpen
  \bibfield  {author} {\bibinfo {author} {\bibfnamefont {R.}~\bibnamefont
  {Winkler}}, \bibinfo {author} {\bibfnamefont {U.}~\bibnamefont {Zulicke}}, \
  and\ \bibinfo {author} {\bibfnamefont {J.}~\bibnamefont {Bolte}},\
  }\href@noop {} {\bibfield  {journal} {\bibinfo  {journal} {Phys. Rev. B}\
  }\textbf {\bibinfo {volume} {75}},\ \bibinfo {pages} {205314} (\bibinfo
  {year} {2006})}\BibitemShut {NoStop}%
\bibitem [{\citenamefont {Rusin}\ and\ \citenamefont
  {Zawadzki}(2009)}]{RusinZawadzki2009}%
  \BibitemOpen
  \bibfield  {author} {\bibinfo {author} {\bibfnamefont {T.~M.}\ \bibnamefont
  {Rusin}}\ and\ \bibinfo {author} {\bibfnamefont {W.}~\bibnamefont
  {Zawadzki}},\ }\href {\doibase 10.1088/0953-8984/21/13/136219} {\bibfield
  {journal} {\bibinfo  {journal} {Journal of Physics: Condensed Matter}\
  }\textbf {\bibinfo {volume} {21}},\ \bibinfo {pages} {136219} (\bibinfo
  {year} {2009})}\BibitemShut {NoStop}%
\bibitem [{\citenamefont {Zawadzki}(2010)}]{Zawadzki2010}%
  \BibitemOpen
  \bibfield  {author} {\bibinfo {author} {\bibfnamefont {W.}~\bibnamefont
  {Zawadzki}},\ }\href {\doibase 10.1016/j.physleta.2010.05.060} {\bibfield
  {journal} {\bibinfo  {journal} {Physics Letters A}\ }\textbf {\bibinfo
  {volume} {374}},\ \bibinfo {pages} {3177} (\bibinfo {year}
  {2010})}\BibitemShut {NoStop}%
\bibitem [{\citenamefont {Hamner}\ \emph {et~al.}(2014)\citenamefont {Hamner},
  \citenamefont {Zhang}, \citenamefont {Khamehchi}, \citenamefont {Davis},\
  and\ \citenamefont {Engels}}]{Hamner2014}%
  \BibitemOpen
  \bibfield  {author} {\bibinfo {author} {\bibfnamefont {C.}~\bibnamefont
  {Hamner}}, \bibinfo {author} {\bibfnamefont {Y.}~\bibnamefont {Zhang}},
  \bibinfo {author} {\bibfnamefont {M.~A.}\ \bibnamefont {Khamehchi}}, \bibinfo
  {author} {\bibfnamefont {M.~J.}\ \bibnamefont {Davis}}, \ and\ \bibinfo
  {author} {\bibfnamefont {P.}~\bibnamefont {Engels}},\ }\href {\doibase
  10.1038/ncomms5023} {\bibfield  {journal} {\bibinfo  {journal} {Nature
  Communications}\ }\textbf {\bibinfo {volume} {5}},\ \bibinfo {pages} {4023}
  (\bibinfo {year} {2014})}\BibitemShut {NoStop}%
\bibitem [{\citenamefont {Luttinger}(1958)}]{Luttinger_AHE_PR58}%
  \BibitemOpen
  \bibfield  {author} {\bibinfo {author} {\bibfnamefont {J.~M.}\ \bibnamefont
  {Luttinger}},\ }\href@noop {} {\bibfield  {journal} {\bibinfo  {journal}
  {Phys. Rev.}\ }\textbf {\bibinfo {volume} {112}},\ \bibinfo {pages} {739}
  (\bibinfo {year} {1958})}\BibitemShut {NoStop}%
\bibitem [{\citenamefont {Vasko}\ and\ \citenamefont
  {Raichev}(2005)}]{Vasko2005}%
  \BibitemOpen
  \bibfield  {author} {\bibinfo {author} {\bibfnamefont {F.~T.}\ \bibnamefont
  {Vasko}}\ and\ \bibinfo {author} {\bibfnamefont {O.~E.}\ \bibnamefont
  {Raichev}},\ }\href@noop {} {\emph {\bibinfo {title} {Quantum Kinetic Theory
  and Applications: Electrons, Photons, Phonons}}},\ \bibinfo {edition} {1st}\
  ed.\ (\bibinfo  {publisher} {Springer-Verlag New York},\ \bibinfo {year}
  {2005})\BibitemShut {NoStop}%
\bibitem [{\citenamefont {Sinitsyn}\ \emph {et~al.}(2006)\citenamefont
  {Sinitsyn}, \citenamefont {Hill}, \citenamefont {Min}, \citenamefont
  {Sinova},\ and\ \citenamefont {MacDonald}}]{Sinitsyn2006_II}%
  \BibitemOpen
  \bibfield  {author} {\bibinfo {author} {\bibfnamefont {N.~A.}\ \bibnamefont
  {Sinitsyn}}, \bibinfo {author} {\bibfnamefont {J.~E.}\ \bibnamefont {Hill}},
  \bibinfo {author} {\bibfnamefont {H.}~\bibnamefont {Min}}, \bibinfo {author}
  {\bibfnamefont {J.}~\bibnamefont {Sinova}}, \ and\ \bibinfo {author}
  {\bibfnamefont {A.~H.}\ \bibnamefont {MacDonald}},\ }\href {\doibase
  10.1103/PhysRevLett.97.106804} {\bibfield  {journal} {\bibinfo  {journal}
  {Phys. Rev. Lett.}\ }\textbf {\bibinfo {volume} {97}},\ \bibinfo {pages}
  {106804} (\bibinfo {year} {2006})}\BibitemShut {NoStop}%
\bibitem [{\citenamefont {Tworzydlo}\ \emph {et~al.}(2006)\citenamefont
  {Tworzydlo}, \citenamefont {Trauzettel}, \citenamefont {Titov}, \citenamefont
  {Rycerz},\ and\ \citenamefont {Beenakker}}]{Tworzydlo2006}%
  \BibitemOpen
  \bibfield  {author} {\bibinfo {author} {\bibfnamefont {J.}~\bibnamefont
  {Tworzydlo}}, \bibinfo {author} {\bibfnamefont {B.}~\bibnamefont
  {Trauzettel}}, \bibinfo {author} {\bibfnamefont {M.}~\bibnamefont {Titov}},
  \bibinfo {author} {\bibfnamefont {A.}~\bibnamefont {Rycerz}}, \ and\ \bibinfo
  {author} {\bibfnamefont {C.~W.~J.}\ \bibnamefont {Beenakker}},\ }\href
  {\doibase 10.1103/PhysRevLett.96.246802} {\bibfield  {journal} {\bibinfo
  {journal} {Phys. Rev. Lett.}\ }\textbf {\bibinfo {volume} {96}},\ \bibinfo
  {pages} {246802} (\bibinfo {year} {2006})}\BibitemShut {NoStop}%
\bibitem [{Note1()}]{Note1}%
  \BibitemOpen
  \bibinfo {note} {The choice of initial time is arbitrary and I have taken it
  to be zero. Replacing that with a finite initial time $t_0$ simply changes
  $t$ to $t-t_0$ in the oscillatory factors, for example $\sin \omega (t-t_0)$,
  without changing the physical conclusions.}\BibitemShut {Stop}%
\bibitem [{\citenamefont {Provost}\ and\ \citenamefont
  {Vall{\'e}e}(1980)}]{Provost1980}%
  \BibitemOpen
  \bibfield  {author} {\bibinfo {author} {\bibfnamefont {J.~P.}\ \bibnamefont
  {Provost}}\ and\ \bibinfo {author} {\bibfnamefont {G.}~\bibnamefont
  {Vall{\'e}e}},\ }\href {\doibase 10.1007/BF02193559} {\bibfield  {journal}
  {\bibinfo  {journal} {Communications in Mathematical Physics}\ }\textbf
  {\bibinfo {volume} {76}},\ \bibinfo {pages} {289} (\bibinfo {year}
  {1980})}\BibitemShut {NoStop}%
\bibitem [{\citenamefont {Resta}(1994)}]{resta-polarization-review}%
  \BibitemOpen
  \bibfield  {author} {\bibinfo {author} {\bibfnamefont {R.}~\bibnamefont
  {Resta}},\ }\href@noop {} {\bibfield  {journal} {\bibinfo  {journal} {Reviews
  of modern physics}\ }\textbf {\bibinfo {volume} {66}},\ \bibinfo {pages}
  {899} (\bibinfo {year} {1994})}\BibitemShut {NoStop}%
\bibitem [{\citenamefont {Culcer}(2012)}]{Culcer2012}%
  \BibitemOpen
  \bibfield  {author} {\bibinfo {author} {\bibfnamefont {D.}~\bibnamefont
  {Culcer}},\ }\href {\doibase https://doi.org/10.1016/j.physe.2011.11.003}
  {\bibfield  {journal} {\bibinfo  {journal} {Physica E: Low-dimensional
  Systems and Nanostructures}\ }\textbf {\bibinfo {volume} {44}},\ \bibinfo
  {pages} {860} (\bibinfo {year} {2012})},\ \bibinfo {note} {sI:Topological
  Insulators}\BibitemShut {NoStop}%
\bibitem [{\citenamefont {Ludwig}\ \emph {et~al.}(1994)\citenamefont {Ludwig},
  \citenamefont {Fisher}, \citenamefont {Shankar},\ and\ \citenamefont
  {Grinstein}}]{Ludwig1994}%
  \BibitemOpen
  \bibfield  {author} {\bibinfo {author} {\bibfnamefont {A.~W.~W.}\
  \bibnamefont {Ludwig}}, \bibinfo {author} {\bibfnamefont {M.~P.~A.}\
  \bibnamefont {Fisher}}, \bibinfo {author} {\bibfnamefont {R.}~\bibnamefont
  {Shankar}}, \ and\ \bibinfo {author} {\bibfnamefont {G.}~\bibnamefont
  {Grinstein}},\ }\href {\doibase 10.1103/PhysRevB.50.7526} {\bibfield
  {journal} {\bibinfo  {journal} {Phys. Rev. B}\ }\textbf {\bibinfo {volume}
  {50}},\ \bibinfo {pages} {7526} (\bibinfo {year} {1994})}\BibitemShut
  {NoStop}%
\bibitem [{\citenamefont {Culcer}\ \emph {et~al.}(2005)\citenamefont {Culcer},
  \citenamefont {Yao},\ and\ \citenamefont {Niu}}]{Culcer-2005-WPK}%
  \BibitemOpen
  \bibfield  {author} {\bibinfo {author} {\bibfnamefont {D.}~\bibnamefont
  {Culcer}}, \bibinfo {author} {\bibfnamefont {Y.}~\bibnamefont {Yao}}, \ and\
  \bibinfo {author} {\bibfnamefont {Q.}~\bibnamefont {Niu}},\ }\href {\doibase
  10.1103/PhysRevB.72.085110} {\bibfield  {journal} {\bibinfo  {journal} {Phys.
  Rev. B}\ }\textbf {\bibinfo {volume} {72}},\ \bibinfo {pages} {085110}
  (\bibinfo {year} {2005})}\BibitemShut {NoStop}%
\end{thebibliography}%

\end{document}